\documentclass[journal=nalefd,manuscript=article]{achemso}
\setkeys{acs}{etalmode=truncate,maxauthors=20}

\usepackage{graphicx,textcomp,natmove}

\author{Mukesh Tripathi}
\author{Andreas Mittelberger}
\affiliation{University of Vienna, Faculty of Physics, 1090 Vienna, Austria}

\author{Nicholas A. Pike}
\affiliation{Centre for Materials Science and Nanotechnology, University of Oslo, NO-0349 Oslo, Norway}
\alsoaffiliation{nanomat/Q-mat/CESAM, Universit\'e de Li\`ege, Institut de Physique, B-4000 Sart Tilman, Li\`ege, Belgium}

\author{Clemens Mangler}
\author{Jannik C. Meyer}
\affiliation{University of Vienna, Faculty of Physics, 1090 Vienna, Austria}

\author{Matthieu J. Verstraete}
\affiliation{nanomat/Q-mat/CESAM, Universit\'e de Li\`ege, Institut de Physique, B-4000 Sart Tilman, Li\`ege, Belgium}

\author{Jani Kotakoski}
\author{Toma Susi}
\affiliation{University of Vienna, Faculty of Physics, 1090 Vienna, Austria}
\email{toma.susi@univie.ac.at}
\phone{+43-1-427772855}

\title{Electron-Beam Manipulation of Silicon Dopants in Graphene}


\begin{document}
\clearpage

\begin{abstract}
The direct manipulation of individual atoms in materials using scanning probe microscopy has been a seminal achievement of nanotechnology. Recent advances in imaging resolution and sample stability have made scanning transmission electron microscopy a promising alternative for single-atom manipulation of covalently bound materials. Pioneering experiments using an atomically focused electron beam have demonstrated the directed movement of silicon atoms over a handful of sites within the graphene lattice. Here, we achieve a much greater degree of control, allowing us to precisely move silicon impurities along an extended path, circulating a single hexagon, or back and forth between the two graphene sublattices. Even with manual operation, our manipulation rate is already comparable to the state-of-the-art in any atomically precise technique. We further explore the influence of electron energy on the manipulation rate, supported by improved theoretical modeling taking into account the vibrations of atoms near the impurities, and implement feedback to detect manipulation events in real time. In addition to atomic-level engineering of its structure and properties, graphene also provides an excellent platform for refining the accuracy of quantitative models and for the development of automated manipulation.
\end{abstract}

\clearpage

Although single-atom manipulation using scanning probe microscopy was established already 25 years ago~\cite{Eigler90N, Crommie93S}, it continues to provide impressive technological advances, such as atomic memory arrays,~\cite{Kalff16NN} as well as insight into physical phenomena including quantum many-body effects~\cite{Drost17NP, Pavlicek17NN}. However, only relatively weakly bound surface atoms far below room temperature can typically~\cite{Fishlock00N} be affected due to the limited interaction energy with the atomically sharp tip. By contrast, the energetic electrons used in (scanning) transmission electron microscopy (STEM) to image atomic structures can transfer up to tens of electron volts to atoms of light elements such as carbon, allowing the breaking and reforming of covalent bonds. In modern STEM instruments, it is further possible to predominantly direct the electron beam at individual atoms~\cite{Krivanek99UM} in low-dimensional materials such as single-layer graphene~\cite{Meyer08NL}.

Silicon heteroatoms, occasionally found in graphene as substitutional impurities~\cite{Zhou12PRL, Ramasse13NL}, have particularly interesting dynamics. In 2014, they were observed to "jump" through the lattice upon 60 keV electron irradiation~\cite{Susi14PRL}, with no damage to the structure. First-principles simulations revealed the mechanism of Si-C bond inversions: each electron has a finite chance to transfer just enough out-of-plane kinetic energy to one C neighbor to cause it to exchange places with the Si~\cite{Susi14PRL}, a rare example of a direct exchange diffusion in a crystalline material~\cite{Yoshida15Book}. In these early findings, the movement was not controlled, but it was clear that this should be possible by purposefully directing the electron irradiation at the desired C neighbor~\cite{Susi14PRL, Susi15FWF}. 

Recently, some of us achieved the first controlled manipulations~\cite{Susi17UM} by iteratively parking the electron beam for 15~s on top of the C neighbor in the direction the Si should move. These experiments had clear limitations: the Si was moved over just a handful of lattice sites, with unintended double jumps. This lack of control was partly due to the lack of feedback: it was not possible to observe structural changes while the electron beam was parked, resulting in excessive dosing. Dyck et al. recently reported similar manipulations~\cite{Dyck17APL}, in this case by irradiating a small sub-scan window centered on the C neighbor. However, directional control was rather poor, likely due to dosing of undesired C sites. In a subsequent study, the same group was able to move several nearby Si over a few lattice sites~\cite{Dyck17arXiv2}.

Here, we demonstrate greatly improved manipulation of incidental Si impurities in commercial monolayer graphene, identified by their scattering $Z$-contrast~\cite{Krivanek10N} and by electron energy loss spectroscopy~\cite{Ramasse13NL} (Fig.~S1). We now reduced the spot irradiation time to 10~s in our Nion UltraSTEM100 instrument operated in near ultra-high vacuum ($10^{-10}$~mbar) at 60 keV, with a beam current of $35\pm10$~pA corresponding to a dose rate of $(2.2\pm0.6)\times$10$^8$~e$^-$s$^{-1}$ in the $\sim$1.1~\AA\ diameter spot. For another series of experiments conducted at 55~keV, we further implemented real-time feedback by connecting a Keithley 2000 multimeter to the medium angle annular dark field (MAADF) detector and reading out the raw detector voltage averaged over 150~ms while the beam was parked on the desired atom. These values were read into the Nion Swift microscope control software~\cite{Swift} and used to trigger the acquisition of new image frames. As control logic, an increased intensity corresponds to a successful manipulation event (Si has taken the place of the C, leading to greater scattering), with a threshold of 10\% change with respect to the cumulative average of the signal chosen as the trigger condition (see Fig.~S2).

\begin{figure}
\includegraphics[width=0.75\textwidth]{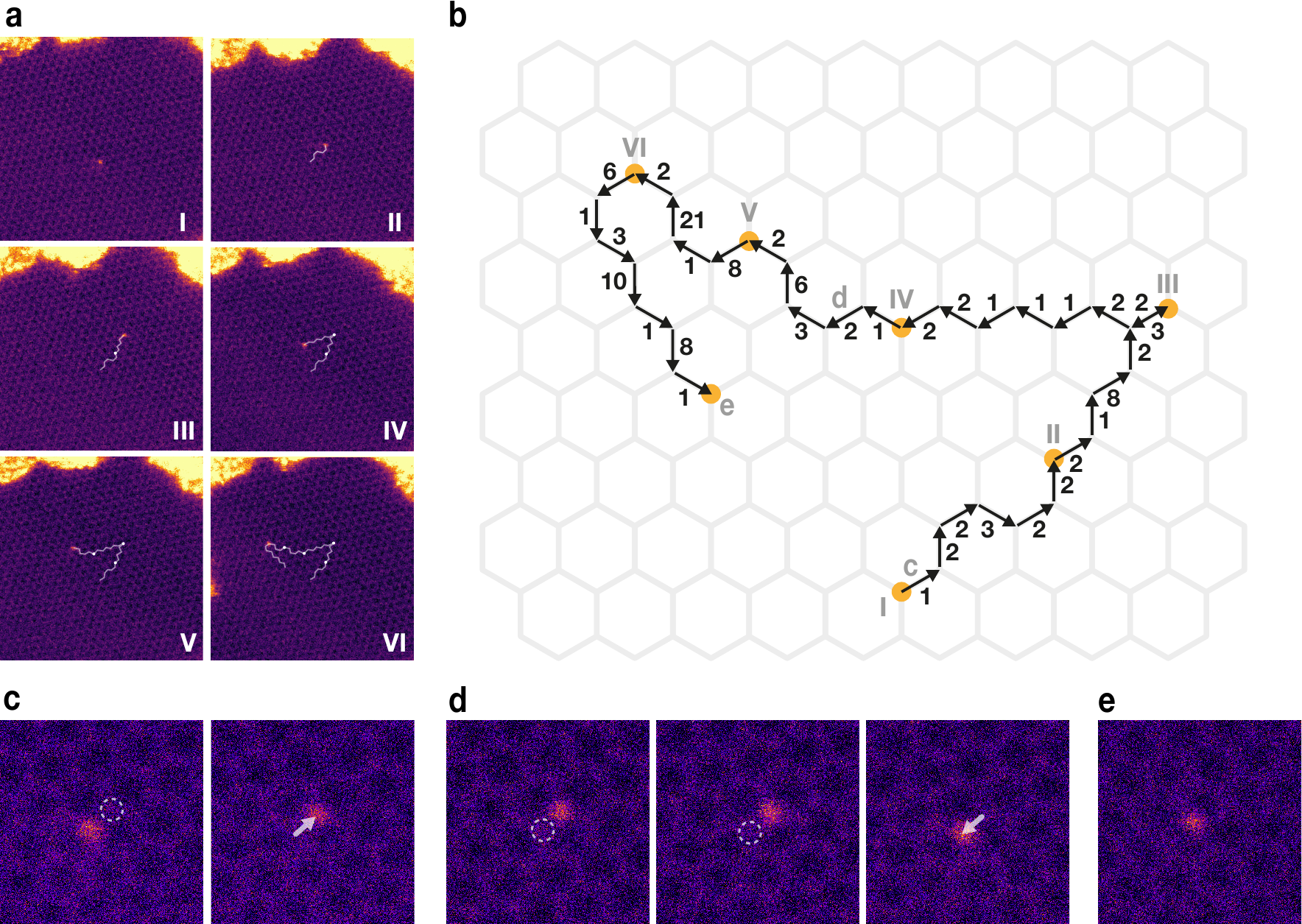}%
\caption{Controlled electron-beam manipulation of a Si heteroatom in graphene (STEM images recorded with a MAADF detector). a) Overviews where the segmented line indicates each of the 34 precisely directed lattice jumps and dots the locations of the Si atom in each previous panel (I-VI). b) Schematic illustration of the path, with orange circles indicating the position of the Si in each overview labeled with Roman numerals (I-VI), and with Arabic numerals indicating the number of 10~s spot irradiations required for each jump. c) Closer views before and after the first jump. The location where the electron beam was parked is indicated by the dashed open circle. d) Closer views of the two frames before and after the 20$^\mathrm{th}$ jump. e) Closer view near the endpoint of the sequence, where a C atom has been knocked out, resulting in fourfold-coordination of the Si.\label{fig:manipulation}}
\end{figure}

\begin{figure}
\includegraphics[width=0.7\textwidth]{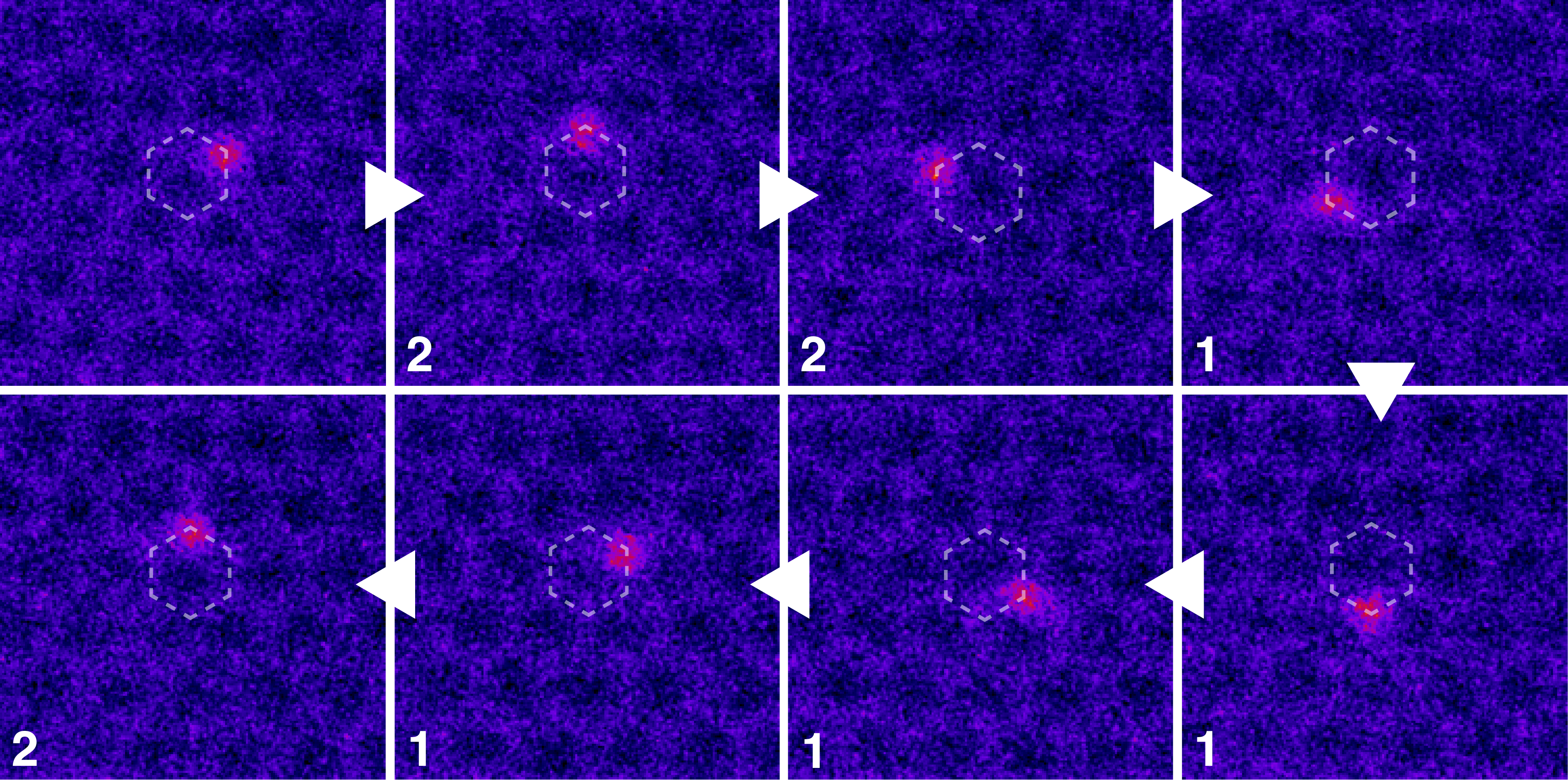}%
\caption{Electron-beam manipulation of a Si heteroatom around a single hexagon in graphene (aligned and colored STEM/MAADF images). The overlaid numbers show the number of 10~s spot irradiations preceding each frame, and the triangles indicate the ordering of the frames.\label{fig:manipulation2}}
\end{figure}

\begin{figure}
\includegraphics[width=0.6\textwidth]{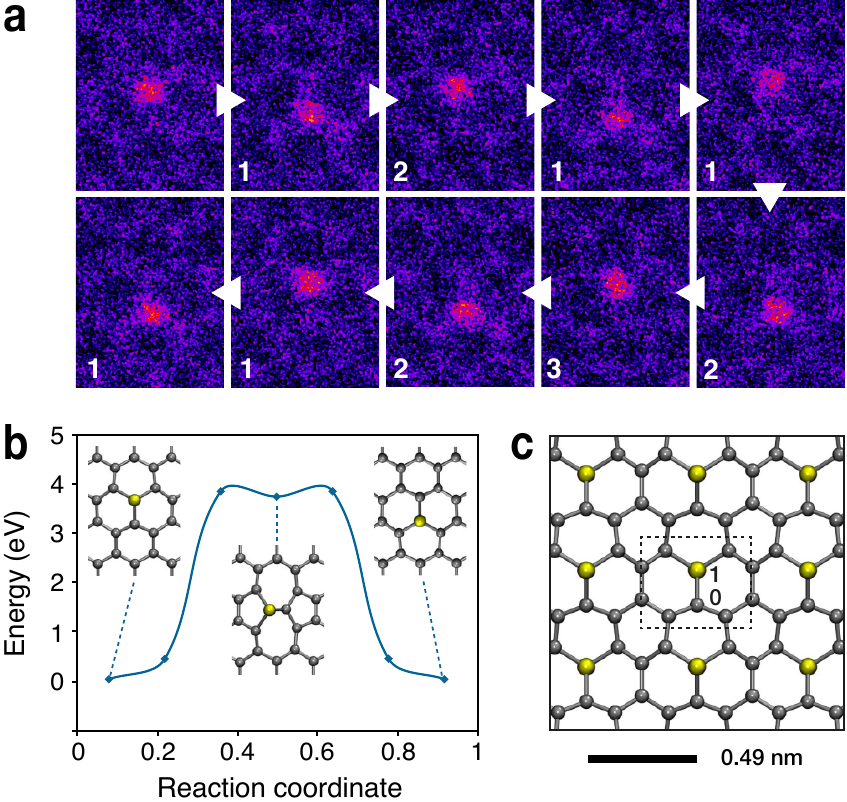}
\caption{Using a Si heteroatom as an atomic bit. a) A single Si heteroatom repeatedly moved from one graphene sub-lattice to the other (aligned and colored STEM/MAADF images). The overlaid numbers show the number of 10~s spot irradiations preceding each frame, and the triangles indicate the ordering of the frames. b) The Si migration barrier calculated with the nudged elastic band method within DFT is close to 4 eV. c) The dash-outlined 4.94$\times$4.28 \AA\ graphene area contains a single Si atom, whose position on either of the sublattices could correspond to a bit value of either 0 or 1.\label{fig:Si-bit}}
\end{figure}

To demonstrate the control we can now achieve, we conducted three types of manipulation experiments at 60~kV. First, we moved a Si atom over a path consisting of 34 lattice jumps precisely in the selected directions, with no undesired motion or double jumps (Fig.~\ref{fig:manipulation}). Second, instead of extended lateral motion, we moved a Si atom around a graphene hexagon (selected frames from a sequence of 75 jumps, which included several double jumps, are shown in Fig.~\ref{fig:manipulation2}); the probability of just the shown motion being a random walk would be $(1/3)^{8} < 0.02$\%. Finally, individual Si atoms could be moved back and forth between the two graphene sublattices (selected frames of 67 jumps are shown in Fig.~\ref{fig:Si-bit}a). At 55~keV, we repeated the latter two kinds of experiments, observing very few double jumps due to the now implemented feedback, and collecting further statistical data on the required doses for our theoretical analyses. As shown by these examples, our level of control is sufficient for creating extended structures~\cite{NosratyAlamdary17PSSB} once the density of impurities can be increased, for example by purposefully implanting them into the lattice~\cite{Bangert13NL, Susi172DM} or by capturing them in vacancies~\cite{Dyck17arXiv2}.

\begin{figure}
\includegraphics[width=0.6\textwidth]{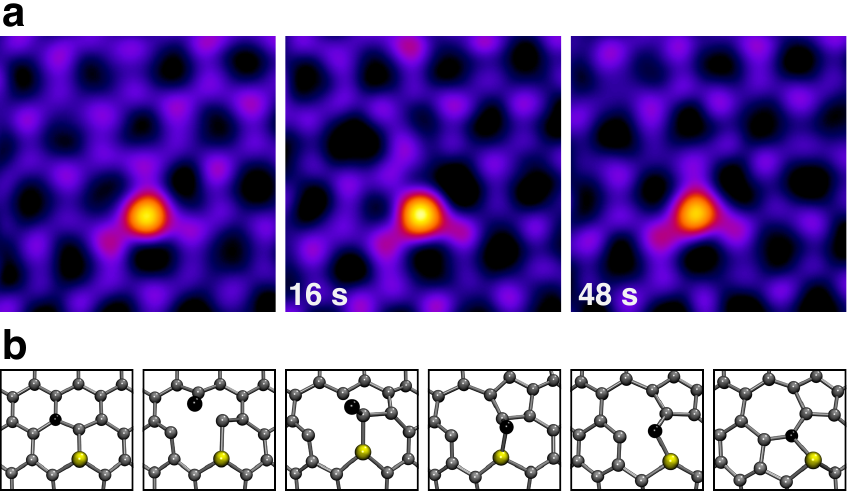}%
\caption{Stone-Wales (SW) transformation near a Si impurity. a) Three consecutive STEM/MAADF frames recorded at 55 kV of a SW transformation induced by the electron beam. The inset numbers indicate the number of seconds of spot irradiation between frames. b) Snapshots of a DFT/MD simulation of a 19.75 eV electron impact on the next-nearest C neighbor to the Si, triggering the SW transformation.\label{fig:SW}}
\end{figure}

\begin{figure}[ht!]
\includegraphics[width=0.65\textwidth]{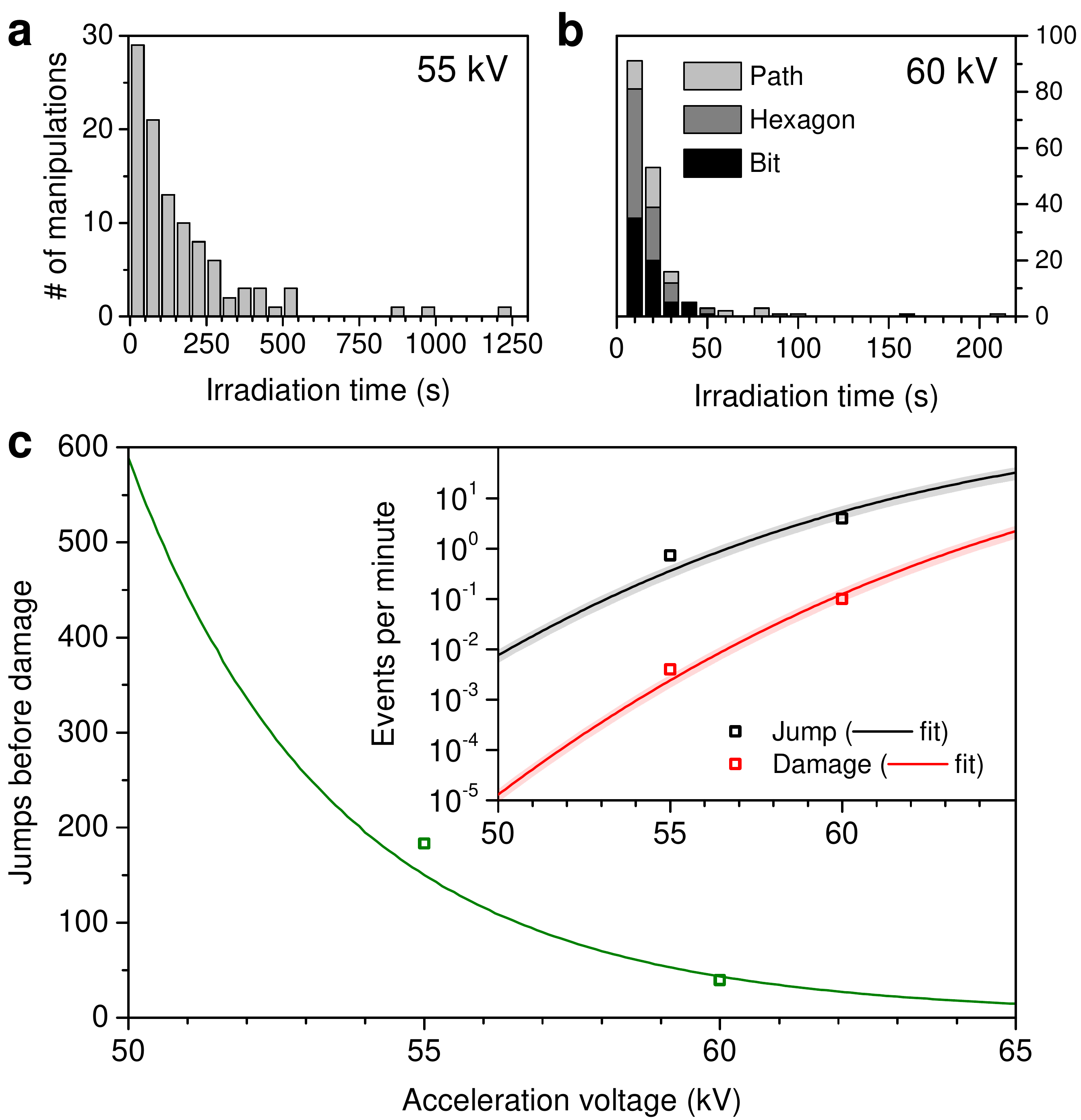}%
\caption{a) Histogram of spot irradiation times for each observed manipulation event at 55~keV (50~s bin width). b) Histogram of 10-second spot irradiations for each manipulation event at 60~keV. c) Theoretically predicted rates of manipulation events (jumps) and knock-on damage, as well as the predicted number of successful manipulations before damage for electron acceleration voltages close to 60~kV and for a realistic beam shape with a current of 35$\pm$10~pA. The open squares correspond to our spot irradiation experiments, the solid lines to our theoretical model with best-fit event threshold energies, and shaded areas to error estimates based on the uncertainty of the irradiation dose.\label{fig:analysis}}
\end{figure}

A stacked histogram of all 60 keV manipulation events (Fig.~\ref{fig:analysis}a) shows that in most cases ($136/164 = 0.83$), the required irradiation time is 20 seconds or less (the geometric mean of 15.1~s providing a good estimate for the Poisson expectation value~\cite{Susi14PRL}), corresponding to a dose of $(8.9\pm0.9) \times 10^8$ electrons per event and a rate close to 4 jumps per minute (neglecting the frame time of 4.2 s). Remarkably, despite the simplicity of the experimental procedure, this is nearly on par with the state-of-the-art in automated scanning tunneling microscopy-based single-atom manipulation~\cite{Kalff16NN}. The two outliers with very long irradiation times of 160 and 210~s are unlikely to follow from the same Poisson process~\cite{Susi14PRL} as the other manipulation events, but are consistent with $^{13}$C atoms (and their frequency, $2/164 = 0.012$, is very close to the natural carbon isotope abundance). This suggests that graphene grown from isotopically purified precursors~\cite{Chen12NM} would be ideal for further large-scale experiments.

The manipulation rate could be increased by increasing the electron acceleration voltage, but this would also increase the probability of knock-on damage. A lower voltage, on the other hand, would decrease the manipulation rate, but also decrease the probability of damage. Due to their differing threshold energies, the damage probability decreases faster than the probability of a jump as a function of electron energy, which means that the statistically expected number of successful jumps can be increased by lowering the voltage. In our experiments at 55 keV, due to our real-time detector feedback, we were also able to obtain more accurate timing for each jump. In total, we moved individual Si impurities 102 times, with an average rate or $0.73\pm0.16$ jumps per minute, corresponding to an average electron irradiation dose per event of $(8.8 \pm 1.0) \times10^{9}$~e$^-$. An example of manipulation around a graphene hexagon is shown in Fig.~S3. In these series, we observed only one event of knock-on damage, with an apparent rate of $0.0040\pm0.0007$ min$^{-1}$ (corresponding to $(8.9 \pm 1.5) \times 10^{11}$~e$^-$). Improving the sharpness of the electron probe would be clearly beneficial: for our expected probe shape~\cite{kotakoski14NC}, we estimate that only 26\% of the dose impinges on the area of the targeted C atom. Thus a sharper probe would allow us to increase the manipulation rate of the atom without impacting the relative probability of damage. 

We can further compare the observed event doses to a first-principles model that includes the effect of atomic vibrations~\cite{Meyer12PRL}. In our previous estimates, we simply used the pristine graphene phonon dispersion~\cite{Susi16NC} even for systems with impurities~\cite{Susi172DMreview}. Here, we have explicitly calculated the phonon dispersion relations for a graphene super-cell containing 71 C atoms and a single Si impurity and compared this to a monolayer of pure graphene. We use density functional perturbation theory as implemented with the \textit{ABINIT} software package~\cite{Gonze2005, Gonze2009, Gonze2016}, with norm-conserving pseudo-potentials generated with the ONCVPSP code~\cite{Hamann2013, Setten2018}, a plane-wave basis set, and the PBE exchange-correlation functional~\cite{Perdew1996} (the differences with LDA in mean-square velocity are a few percent). After calculating the phonon dispersion, a displacement-weighted phonon density of states is constructed for each atom, and populated with Bose-Einstein thermal factors at 300~K (the temperature of our objective lens close to the sample). With harmonic Gaussian modes, the mean-square velocity is then related directly to the mean-square displacement by the mode frequency~\cite{Lee1995}.

With this method, the out-of-plane mean-square velocity $\overline{v^2_z}$ of the C atoms in pristine graphene is around 320 000~m$^2$/s$^2$, as is the mean-square velocity of the nearest-neighbor C atoms to the Si impurity. However, the second nearest neighbors have perceptibly higher velocities, as shown in Fig.~S2. Further neighbor velocities decay exponentially back to the pristine value. We attribute the somewhat surprising `normal' values for the first neighbor velocities to a compensation between (1) softer sp$^3$-like bonding with Si (visible in the vibrational frequencies of C-C single, double and triple bonds), and (2) the mass difference, which for a thermalized harmonic oscillator will increase the velocity of the lighter atom. For second neighbors, the bonding returns to sp$^2$ and the pinning effect of the Si-C block generates the higher velocities.

Interestingly, at 55 keV we observed a new type of a dynamical event: a \mbox{Stone-(Thrower-)} Wales~\cite{Monthioux14C} bond rotation of the next-nearest C-C bond. This process cannot be activated by an elastic electron impact in pristine graphene at such low voltages~\cite{Kotakoski11PRB}, whereas at 60~keV, we believe the back-transformation rate~\cite{Skowron16C} to be too fast to image the defect. However, density functional theory based molecular dynamics (DFT/MD) simulations following our established methodology~\cite{Susi16NC} reveal that the local perturbation caused by the impurity allows this process to be activated by impacts with energies between $[19.675,20.125]$~eV on the next-nearest C neighbor to the Si (and for a perpendicular momentum transfer, contrary to the pristine case~\cite{Kotakoski11PRB}). Combined with the greater velocity of that C atom, this results in a finite event probability even at 55 keV. One of the three such events we observed is shown in Fig.~\ref{fig:SW}, alongside snapshots from an MD simulation elucidating the mechanism.

To compare our model to the manipulation experiments, event cross sections are calculated by integrating the product of a Gaussian velocity distribution $P(v)$ (with width parameter $\overline{v^2_z}$) and the cross section for elastic scattering $\sigma(E_{max}(v,E_e))$ (in the McKinley-Feshbach approximation~\cite{McKinley48PR}) for the maximum kinetic energy $E_{max}(v,E_e)$ that the electron can transfer~\cite{Meyer12PRL} to the nucleus in a momentum and energy conserving collision:
\begin{equation}
\sigma(T,E_e)=\int\! P(v,T) \sigma(E_{max}(v,E_e)) \Theta[E_{max}(v,E_e) - T_d] \, \mathrm{d}v,
\end{equation}
evaluated numerically for all velocities $v$ where the maximum transferred energy exceeds the displacement threshold energy $T_d$ (enforced via the Heaviside step function $\Theta[E_{max}(v,E_e) - T_d]$). $T$ is the temperature (here 300~K) and $E_e$ the electron kinetic energy.

We then find displacement threshold energy values that minimize the least squares error of the theoretical curves with our experimental data, plotted in Fig.~\ref{fig:analysis}c. From the fits, we obtain a threshold of 13.04~eV for the direct exchange process (DFT: 14.8~eV)~\cite{Susi14PRL}, whereas the knock-on threshold is 14.44~eV (DFT: 16.9~eV)~\cite{Susi14PRL}. These discrepancies between experimental and theoretically predicted threshold energies again highlight the inaccuracies of the best available models~\cite{Susi172DMreview}. Since we now include the explicit phonon dispersion for a system with a Si impurity, this shows that the estimate of the displacement threshold energy itself ---essentially a description of dynamical bond breaking--- needs to be improved before accurate quantitative predictions can be made.

Despite the unavoidable chance of knocking out a C atom, STEM manipulation has several advantages over STM. Because of the strong covalent bonding, the Si migration barrier is very large (close to 4~eV, Fig.~\ref{fig:Si-bit}b, as calculated with the nudged elastic band method~\cite{Henkelman00TJoCP} within density functional theory using the GPAW software package~\cite{Enkovaara2010}), making any created arrangements stable at temperatures well above room temperature for extended periods of time. In addition, a beam-stable minimum spacing between the Si impurities might be as low as 5~{\AA} (Fig.~\ref{fig:Si-bit}c). If the position of the Si on either of the two sublattices were to encode a single bit value, this would allow a theoretical density as high as 3000~terabits per square inch, i.e., six times higher than the record achieved by manipulating vacancies in a chlorine surface monolayer using STM~\cite{Kalff16NN}.

Although we emphatically are not proposing that graphene heteroatom bits constitute a practical storage medium in the foreseeable future, there seem to be no fundamental obstacles. At 60~kV, the dose required to cause a jump (flip the bit value) is on the order of $10^9$ e$^-$, whereas the dose falling on the C neighbors of the Si required to record high-quality images is at least two orders of magnitude lower. To simply detect the position of the Si on either of the two sublattices, likely an additional two orders of magnitude lower dose would be sufficient, especially if combined with compressed sensing and computer vision. At 55 kV, due to the greater dose required to move the Si, this difference is one order of magnitude greater still, further reducing the probability of unintended bit  flips. It thus appears feasible to both reliably write and read bit values using the same STEM instrument. The manipulation of other heteroatoms~\cite{Susi172DMreview} along with instrumental advances in beam shaping combined with an optimized acceleration voltage might further improve these capacities.

In conclusion, we have shown how targeted electron irradiation with an \AA ngstr\"om-sized electron probe can reliably move Si impurities through the graphene lattice. The improved degree of control reported here demonstrates the capabilities of scanning transmission electron microscopy as a capable tool for single-atom manipulation in covalently bound materials. Although all atoms in graphene are surface atoms, its impurities are much stronger bound than the surface adatoms or vacancies that can be manipulated with scanning probe techniques, resulting in inherent temperature stability. Furthermore, very recent experiments on manipulating impurities within bulk silicon have provided the first evidence for the applicability of this technique also in three-dimensional crystals~\cite{Hudak18AN}. Nonetheless, graphene remains highly suited for refining the accuracy of quantitative models and the development of automated manipulation. Optimization of the feedback parameters and the implementation of drift compensation will enhance our capabilities further, to be followed by structure recognition~\cite{Ziatdinov17AN} and software-controlled beam re-positioning as the next steps towards full automation.

\section*{Supporting Information} 
Figures S1-S4: EELS spectrum of a Si substitution, MAADF detector feedback time series, example manipulation sequence at 55 keV, and the calculated out-of-plane mean-square velocities of the graphene supercell with a Si impurity. Unprocessed electron microscopy image stacks of the manipulation experiments are provided as open data in Ref.~\citenum{opendata}.

\begin{acknowledgement}
T.S. and M.T. acknowledge the Austrian Science Fund (FWF) project P 28322-N36 for funding. T.S. was also supported by the European Research Council (ERC) Grant No. 756277-ATMEN, and acknowledges the Vienna Scientific Cluster for computer time. A.M., C.M., and J.C.M. were supported by the ERC Grant No. 336453-PICOMAT. J.K. was supported by the FWF project I 3181-N36 and the Wiener Wissenschafts\mbox{-,} Forschungs- und Technologiefonds (WWTF) project MA14-009. N.A.P was supported by the Research Council of Norway through the Frinatek program and both N.A.P and M.J.V. are supported by the Belgian Fonds National de la Recherche Scientifique (FNRS) under grant number PDR T.1077.15-1/7, and acknowledge CECI (FNRS G.A. 2.5020.11) and CENAERO-zenobe (Waloon region G.A. 1117545) for computer time.
\end{acknowledgement}


\begin{mcitethebibliography}{42}
\providecommand*\natexlab[1]{#1}
\providecommand*\mciteSetBstSublistMode[1]{}
\providecommand*\mciteSetBstMaxWidthForm[2]{}
\providecommand*\mciteBstWouldAddEndPuncttrue
  {\def\EndOfBibitem{\unskip.}}
\providecommand*\mciteBstWouldAddEndPunctfalse
  {\let\EndOfBibitem\relax}
\providecommand*\mciteSetBstMidEndSepPunct[3]{}
\providecommand*\mciteSetBstSublistLabelBeginEnd[3]{}
\providecommand*\EndOfBibitem{}
\mciteSetBstSublistMode{f}
\mciteSetBstMaxWidthForm{subitem}{(\alph{mcitesubitemcount})}
\mciteSetBstSublistLabelBeginEnd
  {\mcitemaxwidthsubitemform\space}
  {\relax}
  {\relax}

\bibitem[Eigler and Schweizer(1990)Eigler, and Schweizer]{Eigler90N}
Eigler,~D.~M.; Schweizer,~E.~K. \emph{Nature} \textbf{1990}, \emph{344},
  524--526\relax
\mciteBstWouldAddEndPuncttrue
\mciteSetBstMidEndSepPunct{\mcitedefaultmidpunct}
{\mcitedefaultendpunct}{\mcitedefaultseppunct}\relax
\EndOfBibitem
\bibitem[Crommie \latin{et~al.}(1993)Crommie, Lutz, and Eigler]{Crommie93S}
Crommie,~M.~F.; Lutz,~C.~P.; Eigler,~D.~M. \emph{Science} \textbf{1993},
  \emph{262}, 218--220\relax
\mciteBstWouldAddEndPuncttrue
\mciteSetBstMidEndSepPunct{\mcitedefaultmidpunct}
{\mcitedefaultendpunct}{\mcitedefaultseppunct}\relax
\EndOfBibitem
\bibitem[Kalff \latin{et~al.}(2016)Kalff, Rebergen, Fahrenfort, Girovsky,
  Toskovic, Lado, Fern{\'a}ndez-Rossier, and Otte]{Kalff16NN}
Kalff,~F.~E.; Rebergen,~M.~P.; Fahrenfort,~E.; Girovsky,~J.; Toskovic,~R.;
  Lado,~J.~L.; Fern{\'a}ndez-Rossier,~J.; Otte,~A.~F. \emph{Nat.
  Nanotechnol.} \textbf{2016}, \emph{11}, 926--929\relax
\mciteBstWouldAddEndPuncttrue
\mciteSetBstMidEndSepPunct{\mcitedefaultmidpunct}
{\mcitedefaultendpunct}{\mcitedefaultseppunct}\relax
\EndOfBibitem
\bibitem[Drost \latin{et~al.}(2017)Drost, Ojanen, Harju, and
  Liljeroth]{Drost17NP}
Drost,~R.; Ojanen,~T.; Harju,~A.; Liljeroth,~P. \emph{Nat. Phys.}
  \textbf{2017}, \emph{13}, 668--671\relax
\mciteBstWouldAddEndPuncttrue
\mciteSetBstMidEndSepPunct{\mcitedefaultmidpunct}
{\mcitedefaultendpunct}{\mcitedefaultseppunct}\relax
\EndOfBibitem
\bibitem[Pavli{\v c}ek \latin{et~al.}(2017)Pavli{\v c}ek, Mistry, Majzik, Moll,
  Meyer, Fox, and Gross]{Pavlicek17NN}
Pavli{\v c}ek,~N.; Mistry,~A.; Majzik,~Z.; Moll,~N.; Meyer,~G.; Fox,~D.~J.;
  Gross,~L. \emph{Nat. Nanotechnol.} \textbf{2017}, \emph{12},
  308--311\relax
\mciteBstWouldAddEndPuncttrue
\mciteSetBstMidEndSepPunct{\mcitedefaultmidpunct}
{\mcitedefaultendpunct}{\mcitedefaultseppunct}\relax
\EndOfBibitem
\bibitem[Fishlock \latin{et~al.}(2000)Fishlock, Oral, Egdell, and
  Pethica]{Fishlock00N}
Fishlock,~T.~W.; Oral,~A.; Egdell,~R.~G.; Pethica,~J.~B. \emph{Nature}
  \textbf{2000}, \emph{404}, 743--745\relax
\mciteBstWouldAddEndPuncttrue
\mciteSetBstMidEndSepPunct{\mcitedefaultmidpunct}
{\mcitedefaultendpunct}{\mcitedefaultseppunct}\relax
\EndOfBibitem
\bibitem[Krivanek \latin{et~al.}(1999)Krivanek, Dellby, and
  Lupini]{Krivanek99UM}
Krivanek,~O.~L.; Dellby,~N.; Lupini,~A.~R. \emph{Ultramicroscopy}
  \textbf{1999}, \emph{78}, 1 -- 11\relax
\mciteBstWouldAddEndPuncttrue
\mciteSetBstMidEndSepPunct{\mcitedefaultmidpunct}
{\mcitedefaultendpunct}{\mcitedefaultseppunct}\relax
\EndOfBibitem
\bibitem[Meyer \latin{et~al.}(2008)Meyer, Kisielowski, Erni, Rossell, Crommie,
  and Zettl]{Meyer08NL}
Meyer,~J.~C.; Kisielowski,~C.; Erni,~R.; Rossell,~M.~D.; Crommie,~M.~F.;
  Zettl,~A. \emph{Nano Lett.} \textbf{2008}, \emph{8}, 3582--3586\relax
\mciteBstWouldAddEndPuncttrue
\mciteSetBstMidEndSepPunct{\mcitedefaultmidpunct}
{\mcitedefaultendpunct}{\mcitedefaultseppunct}\relax
\EndOfBibitem
\bibitem[Zhou \latin{et~al.}(2012)Zhou, Kapetanakis, Prange, Pantelides,
  Pennycook, and Idrobo]{Zhou12PRL}
Zhou,~W.; Kapetanakis,~M.; Prange,~M.; Pantelides,~S.; Pennycook,~S.;
  Idrobo,~J.-C. \emph{Phys. Rev. Lett.} \textbf{2012}, \emph{109}, 206803\relax
\mciteBstWouldAddEndPuncttrue
\mciteSetBstMidEndSepPunct{\mcitedefaultmidpunct}
{\mcitedefaultendpunct}{\mcitedefaultseppunct}\relax
\EndOfBibitem
\bibitem[Ramasse \latin{et~al.}(2013)Ramasse, Seabourne, Kepaptsoglou, Zan,
  Bangert, and Scott]{Ramasse13NL}
Ramasse,~Q.~M.; Seabourne,~C.~R.; Kepaptsoglou,~D.-M.; Zan,~R.; Bangert,~U.;
  Scott,~A.~J. \emph{Nano Lett.} \textbf{2013}, \emph{13}, 4989--4995\relax
\mciteBstWouldAddEndPuncttrue
\mciteSetBstMidEndSepPunct{\mcitedefaultmidpunct}
{\mcitedefaultendpunct}{\mcitedefaultseppunct}\relax
\EndOfBibitem
\bibitem[Susi \latin{et~al.}(2014)Susi, Kotakoski, Kepaptsoglou, Mangler,
  Lovejoy, Krivanek, Zan, Bangert, Ayala, Meyer, and Ramasse]{Susi14PRL}
Susi,~T.; Kotakoski,~J.; Kepaptsoglou,~D.; Mangler,~C.; Lovejoy,~T.~C.;
  Krivanek,~O.~L.; Zan,~R.; Bangert,~U.; Ayala,~P.; Meyer,~J.~C.; Ramasse,~Q.
  \emph{Phys. Rev. Lett.} \textbf{2014}, \emph{113}, 115501\relax
\mciteBstWouldAddEndPuncttrue
\mciteSetBstMidEndSepPunct{\mcitedefaultmidpunct}
{\mcitedefaultendpunct}{\mcitedefaultseppunct}\relax
\EndOfBibitem
\bibitem[Yoshida and Langouche(2015)Yoshida, and Langouche]{Yoshida15Book}
Yoshida,~Y.; Langouche,~G. \emph{Defects and Impurities in Silicon Materials:
  An Introduction to Atomic-Level Silicon Engineering}; Lecture Notes in
  Physics; Springer Japan, 2015; Vol. 916\relax
\mciteBstWouldAddEndPuncttrue
\mciteSetBstMidEndSepPunct{\mcitedefaultmidpunct}
{\mcitedefaultendpunct}{\mcitedefaultseppunct}\relax
\EndOfBibitem
\bibitem[Susi(2015)]{Susi15FWF}
Susi,~T. \emph{Res. Ideas Outcomes} \textbf{2015}, \emph{1},
  e7479\relax
\mciteBstWouldAddEndPuncttrue
\mciteSetBstMidEndSepPunct{\mcitedefaultmidpunct}
{\mcitedefaultendpunct}{\mcitedefaultseppunct}\relax
\EndOfBibitem
\bibitem[Susi \latin{et~al.}(2017)Susi, Meyer, and Kotakoski]{Susi17UM}
Susi,~T.; Meyer,~J.; Kotakoski,~J. \emph{Ultramicroscopy} \textbf{2017},
  \emph{180}, 163--172\relax
\mciteBstWouldAddEndPuncttrue
\mciteSetBstMidEndSepPunct{\mcitedefaultmidpunct}
{\mcitedefaultendpunct}{\mcitedefaultseppunct}\relax
\EndOfBibitem
\bibitem[Dyck \latin{et~al.}(2017)Dyck, Kim, Kalinin, and Jesse]{Dyck17APL}
Dyck,~O.; Kim,~S.; Kalinin,~S.~V.; Jesse,~S. \emph{Appl. Phys. Lett.}
  \textbf{2017}, \emph{111}, 113104\relax
\mciteBstWouldAddEndPuncttrue
\mciteSetBstMidEndSepPunct{\mcitedefaultmidpunct}
{\mcitedefaultendpunct}{\mcitedefaultseppunct}\relax
\EndOfBibitem
\bibitem[{Dyck} \latin{et~al.}(2017){Dyck}, {Kim}, {Jimenez-Izal},
  {Alexandrova}, {Kalinin}, and {Jesse}]{Dyck17arXiv2}
{Dyck},~O.; {Kim},~S.; {Jimenez-Izal},~E.; {Alexandrova},~A.~N.;
  {Kalinin},~S.~V.; {Jesse},~S. \emph{arXiv [cond-mat.mtrl-sci]} \textbf{2017}, 1710.09416\relax
\mciteBstWouldAddEndPunctfalse
\mciteSetBstMidEndSepPunct{\mcitedefaultmidpunct}
{}{\mcitedefaultseppunct}\relax
\EndOfBibitem
\bibitem[Krivanek \latin{et~al.}(2010)Krivanek, Chisholm, Nicolosi, Pennycook,
  Corbin, Dellby, Murfitt, Own, Szilagyi, Oxley, Pantelides, and
  Pennycook]{Krivanek10N}
Krivanek,~O.~L.; Chisholm,~M.~F.; Nicolosi,~V.; Pennycook,~T.~J.;
  Corbin,~G.~J.; Dellby,~N.; Murfitt,~M.~F.; Own,~C.~S.; Szilagyi,~Z.~S.;
  Oxley,~M.~P.; Pantelides,~S.~T.; Pennycook,~S.~J. \emph{Nature}
  \textbf{2010}, \emph{464}, 571--574\relax
\mciteBstWouldAddEndPuncttrue
\mciteSetBstMidEndSepPunct{\mcitedefaultmidpunct}
{\mcitedefaultendpunct}{\mcitedefaultseppunct}\relax
\EndOfBibitem
\bibitem[{Nion Co.}(2018)]{Swift}
{Nion Co.}, {Nion Swift User's Guide}, \textbf{2018}, {http://nionswift.readthedocs.io} (accessed June 14, 2018)\relax
\mciteBstWouldAddEndPuncttrue
\mciteSetBstMidEndSepPunct{\mcitedefaultmidpunct}
{\mcitedefaultendpunct}{\mcitedefaultseppunct}\relax
\EndOfBibitem
\bibitem[Nosraty~Alamdary \latin{et~al.}(2017)Nosraty~Alamdary, Kotakoski, and
  Susi]{NosratyAlamdary17PSSB}
Nosraty~Alamdary,~D.; Kotakoski,~J.; Susi,~T. \emph{Phys. Status Solidi B}
  \textbf{2017}, \emph{254}, 1700188\relax
\mciteBstWouldAddEndPuncttrue
\mciteSetBstMidEndSepPunct{\mcitedefaultmidpunct}
{\mcitedefaultendpunct}{\mcitedefaultseppunct}\relax
\EndOfBibitem
\bibitem[Bangert \latin{et~al.}(2013)Bangert, Pierce, Kepaptsoglou, Ramasse,
  Zan, Gass, Van~den Berg, Boothroyd, Amani, and Hofs{\"a}ss]{Bangert13NL}
Bangert,~U.; Pierce,~W.; Kepaptsoglou,~D.~M.; Ramasse,~Q.; Zan,~R.;
  Gass,~M.~H.; Van~den Berg,~J.~A.; Boothroyd,~C.~B.; Amani,~J.;
  Hofs{\"a}ss,~H. \emph{Nano Lett.} \textbf{2013}, \emph{13},
  4902--4907\relax
\mciteBstWouldAddEndPuncttrue
\mciteSetBstMidEndSepPunct{\mcitedefaultmidpunct}
{\mcitedefaultendpunct}{\mcitedefaultseppunct}\relax
\EndOfBibitem
\bibitem[Susi \latin{et~al.}(2017)Susi, Hardcastle, Hofs{\"a}ss, Mittelberger,
  Pennycook, Mangler, Drummond-Brydson, Scott, Meyer, and Kotakoski]{Susi172DM}
Susi,~T.; Hardcastle,~T.~P.; Hofs{\"a}ss,~H.; Mittelberger,~A.;
  Pennycook,~T.~J.; Mangler,~C.; Drummond-Brydson,~R.; Scott,~A.~J.;
  Meyer,~J.~C.; Kotakoski,~J. \emph{2D Mater.} \textbf{2017}, \emph{4},
  021013\relax
\mciteBstWouldAddEndPuncttrue
\mciteSetBstMidEndSepPunct{\mcitedefaultmidpunct}
{\mcitedefaultendpunct}{\mcitedefaultseppunct}\relax
\EndOfBibitem
\bibitem[Chen \latin{et~al.}(2012)Chen, Wu, Mishra, Kang, Zhang, Cho, Cai,
  Balandin, and Ruoff]{Chen12NM}
Chen,~S.; Wu,~Q.; Mishra,~C.; Kang,~J.; Zhang,~H.; Cho,~K.; Cai,~W.;
  Balandin,~A.~A.; Ruoff,~R.~S. \emph{Nat. Mater.} \textbf{2012},
  \emph{11}, 203--207\relax
\mciteBstWouldAddEndPuncttrue
\mciteSetBstMidEndSepPunct{\mcitedefaultmidpunct}
{\mcitedefaultendpunct}{\mcitedefaultseppunct}\relax
\EndOfBibitem
\bibitem[Kotakoski \latin{et~al.}(2014)Kotakoski, Mangler, and
  Meyer]{kotakoski14NC}
Kotakoski,~J.; Mangler,~C.; Meyer,~J.~C. \emph{Nat. Commun.}
  \textbf{2014}, \emph{5}, 3991\relax
\mciteBstWouldAddEndPuncttrue
\mciteSetBstMidEndSepPunct{\mcitedefaultmidpunct}
{\mcitedefaultendpunct}{\mcitedefaultseppunct}\relax
\EndOfBibitem
\bibitem[Meyer \latin{et~al.}(2012)Meyer, Eder, Kurasch, Skakalova, Kotakoski,
  Park, Roth, Chuvilin, Eyhusen, Benner, Krasheninnikov, and
  Kaiser]{Meyer12PRL}
Meyer,~J.~C.; Eder,~F.; Kurasch,~S.; Skakalova,~V.; Kotakoski,~J.; Park,~H.~J.;
  Roth,~S.; Chuvilin,~A.; Eyhusen,~S.; Benner,~G.; Krasheninnikov,~A.~V.;
  Kaiser,~U. \emph{Phys. Rev. Lett.} \textbf{2012}, \emph{108}, 196102\relax
\mciteBstWouldAddEndPuncttrue
\mciteSetBstMidEndSepPunct{\mcitedefaultmidpunct}
{\mcitedefaultendpunct}{\mcitedefaultseppunct}\relax
\EndOfBibitem
\bibitem[Susi \latin{et~al.}(2016)Susi, Hofer, Argentero, Leuthner, Pennycook,
  Mangler, Meyer, and Kotakoski]{Susi16NC}
Susi,~T.; Hofer,~C.; Argentero,~G.; Leuthner,~G.~T.; Pennycook,~T.~J.;
  Mangler,~C.; Meyer,~J.~C.; Kotakoski,~J. \emph{Nat. Commun.}
  \textbf{2016}, \emph{7}, 13040\relax
\mciteBstWouldAddEndPuncttrue
\mciteSetBstMidEndSepPunct{\mcitedefaultmidpunct}
{\mcitedefaultendpunct}{\mcitedefaultseppunct}\relax
\EndOfBibitem
\bibitem[Susi \latin{et~al.}(2017)Susi, Kepaptsoglou, Lin, Ramasse, Meyer,
  Suenaga, and Kotakoski]{Susi172DMreview}
Susi,~T.; Kepaptsoglou,~D.; Lin,~Y.-C.; Ramasse,~Q.; Meyer,~J.~C.; Suenaga,~K.;
  Kotakoski,~J. \emph{2D Mater.} \textbf{2017}, \emph{4}\relax
\mciteBstWouldAddEndPuncttrue
\mciteSetBstMidEndSepPunct{\mcitedefaultmidpunct}
{\mcitedefaultendpunct}{\mcitedefaultseppunct}\relax
\EndOfBibitem
\bibitem[Gonze \latin{et~al.}(2005)Gonze, Rignanese, Verstraete, Beuken,
  Pouillon, Caracas, Jollet, Torrent, Zerah, Mikami, Ghosez, Veithen, Raty,
  Olevano, Bruneval, Reining, Godby, Onida, Hamann, and Allan]{Gonze2005}
Gonze,~X.; Rignanese,~G.-M.; Verstraete,~M.; Beuken,~J.-M.; Pouillon,~Y.;
  Caracas,~R.; Jollet,~F.; Torrent,~M.; Zerah,~G.; Mikami,~M.; Ghosez,~P.;
  Veithen,~M.; Raty,~J.-Y.; Olevano,~V.; Bruneval,~F.; Reining,~L.; Godby,~R.;
  Onida,~G.; Hamann,~D.; Allan,~D. \emph{Z. Kristallogr.} \textbf{2005},
  \emph{220}, 558--562\relax
\mciteBstWouldAddEndPuncttrue
\mciteSetBstMidEndSepPunct{\mcitedefaultmidpunct}
{\mcitedefaultendpunct}{\mcitedefaultseppunct}\relax
\EndOfBibitem
\bibitem[Gonze \latin{et~al.}(2009)Gonze, Amadon, Anglade, Beuken, Bottin,
  Boulanger, Bruneval, Caliste, Caracas, Cote, Deutsch, Genovese, Ghosez,
  Giantomassi, Goedecker, Hamann, Hermet, Jollet, Jomard, Leroux, Mancini,
  Mazevet, Oliveira, Onida, Pouillon, Rangel, Rignanese, Sangalli, Shaltaf,
  Torrent, Verstraete, Z{\'e}rah, and Zwanziger]{Gonze2009}
Gonze,~X.; Amadon,~B.; Anglade,~P.~M.; Beuken,~J.-M.; Bottin,~F.;
  Boulanger,~P.; Bruneval,~F.; Caliste,~D.; Caracas,~R.; Cote,~M.; Deutsch,~T.;
  Genovese,~L.; Ghosez,~P.; Giantomassi,~M.; Goedecker,~S.; Hamann,~D.;
  Hermet,~P.; Jollet,~F.; Jomard,~G.; Leroux,~S. \latin{et~al.}  \emph{Comp.
  Phys. Comm.} \textbf{2009}, \emph{180}, 2582--2612\relax
\mciteBstWouldAddEndPuncttrue
\mciteSetBstMidEndSepPunct{\mcitedefaultmidpunct}
{\mcitedefaultendpunct}{\mcitedefaultseppunct}\relax
\EndOfBibitem
\bibitem[Gonze \latin{et~al.}(2016)Gonze, Jollet, Araujo, Adams, Amadon,
  T.Applencourt, Audouze, Beuken, Bieder, Bokhanchuk, Bousquet, Bruneval,
  Caliste, Cote, Dahm, Pieve, Delaveau, Gennaro, Dorado, C.Espejo, Geneste,
  Genovese, Gerossier, Giantomassi, Gillet, Hamann, L.He, Jomard, Janssen,
  Roux, Levitt, Lherbier, Liu, Lukacevic, Martin, Martins, Oliveira, Ponce,
  Pouillon, Rangel, Rignanese, Romero, Rousseau, Rubel, Shukri, Stankovski,
  Torrent, VanSetten, Troeye, Verstraete, Waroquiers, Wiktor, Xu, Zhou, and
  Zwanziger]{Gonze2016}
Gonze,~X.; Jollet,~F.; Araujo,~F.~A.; Adams,~D.; Amadon,~B.; T.Applencourt,;
  Audouze,~C.; Beuken,~J.-M.; Bieder,~J.; Bokhanchuk,~A.; Bousquet,~E.;
  Bruneval,~F.; Caliste,~D.; Cote,~M.; Dahm,~F.; Pieve,~F.~D.; Delaveau,~M.;
  Gennaro,~M.~D.; Dorado,~B.; C.Espejo, \latin{et~al.}  \emph{Comp. Phys.
  Comm.} \textbf{2016}, \emph{205}, 106--131\relax
\mciteBstWouldAddEndPuncttrue
\mciteSetBstMidEndSepPunct{\mcitedefaultmidpunct}
{\mcitedefaultendpunct}{\mcitedefaultseppunct}\relax
\EndOfBibitem
\bibitem[Hamann(2013)]{Hamann2013}
Hamann,~D.~R. \emph{Phys. Rev. B} \textbf{2013}, \emph{88}, 085117\relax
\mciteBstWouldAddEndPuncttrue
\mciteSetBstMidEndSepPunct{\mcitedefaultmidpunct}
{\mcitedefaultendpunct}{\mcitedefaultseppunct}\relax
\EndOfBibitem
\bibitem[van Setten \latin{et~al.}(2018)van Setten, Giantomassi, Bousquet,
  Verstraete, Hamann, Gonze, and Rignanese]{Setten2018}
van Setten,~M.; Giantomassi,~M.; Bousquet,~E.; Verstraete,~M.; Hamann,~D.;
  Gonze,~X.; Rignanese,~G.-M. \emph{Comput. Phys. Commun.}
  \textbf{2018}, \emph{226}, 39\relax
\mciteBstWouldAddEndPuncttrue
\mciteSetBstMidEndSepPunct{\mcitedefaultmidpunct}
{\mcitedefaultendpunct}{\mcitedefaultseppunct}\relax
\EndOfBibitem
\bibitem[Perdew \latin{et~al.}(1996)Perdew, Burke, and Ernzerhof]{Perdew1996}
Perdew,~J.~P.; Burke,~K.; Ernzerhof,~M. \emph{Phys. Rev. Lett.} \textbf{1996},
  \emph{77}, 3865\relax
\mciteBstWouldAddEndPuncttrue
\mciteSetBstMidEndSepPunct{\mcitedefaultmidpunct}
{\mcitedefaultendpunct}{\mcitedefaultseppunct}\relax
\EndOfBibitem
\bibitem[Lee and Gonze(1995)Lee, and Gonze]{Lee1995}
Lee,~C.; Gonze,~X. \emph{Phys. Rev. B} \textbf{1995}, \emph{51}, 8610\relax
\mciteBstWouldAddEndPuncttrue
\mciteSetBstMidEndSepPunct{\mcitedefaultmidpunct}
{\mcitedefaultendpunct}{\mcitedefaultseppunct}\relax
\EndOfBibitem
\bibitem[Monthioux and Charlier(2014)Monthioux, and Charlier]{Monthioux14C}
Monthioux,~M.; Charlier,~J.-C. \emph{Carbon} \textbf{2014}, \emph{75}, 1 --
  4\relax
\mciteBstWouldAddEndPuncttrue
\mciteSetBstMidEndSepPunct{\mcitedefaultmidpunct}
{\mcitedefaultendpunct}{\mcitedefaultseppunct}\relax
\EndOfBibitem
\bibitem[Kotakoski \latin{et~al.}(2011)Kotakoski, Meyer, Kurasch,
  Santos-Cottin, Kaiser, and Krasheninnikov]{Kotakoski11PRB}
Kotakoski,~J.; Meyer,~J.~C.; Kurasch,~S.; Santos-Cottin,~D.; Kaiser,~U.;
  Krasheninnikov,~A.~V. \emph{Phys. Rev. B} \textbf{2011}, \emph{83},
  245420\relax
\mciteBstWouldAddEndPuncttrue
\mciteSetBstMidEndSepPunct{\mcitedefaultmidpunct}
{\mcitedefaultendpunct}{\mcitedefaultseppunct}\relax
\EndOfBibitem
\bibitem[Skowron \latin{et~al.}(2016)Skowron, Koroteev, Baldoni, Lopatin,
  Zurutuza, Chuvilin, and Besley]{Skowron16C}
Skowron,~S.~T.; Koroteev,~V.~O.; Baldoni,~M.; Lopatin,~S.; Zurutuza,~A.;
  Chuvilin,~A.; Besley,~E. \emph{Carbon} \textbf{2016}, \emph{105},
  176--182\relax
\mciteBstWouldAddEndPuncttrue
\mciteSetBstMidEndSepPunct{\mcitedefaultmidpunct}
{\mcitedefaultendpunct}{\mcitedefaultseppunct}\relax
\EndOfBibitem
\bibitem[McKinley and Feshbach(1948)McKinley, and Feshbach]{McKinley48PR}
McKinley,~W.~A.,~Jr.; Feshbach,~H. \emph{Phys. Rev.} \textbf{1948}, \emph{74},
  1759--1763\relax
\mciteBstWouldAddEndPuncttrue
\mciteSetBstMidEndSepPunct{\mcitedefaultmidpunct}
{\mcitedefaultendpunct}{\mcitedefaultseppunct}\relax
\EndOfBibitem
\bibitem[Henkelman and J\'onsson(2000)Henkelman, and
  J\'onsson]{Henkelman00TJoCP}
Henkelman,~G.; J\'onsson,~H. \emph{J. Chem. Phys.} \textbf{2000}, \emph{113},
  9978\relax
\mciteBstWouldAddEndPuncttrue
\mciteSetBstMidEndSepPunct{\mcitedefaultmidpunct}
{\mcitedefaultendpunct}{\mcitedefaultseppunct}\relax
\EndOfBibitem
\bibitem[Enkovaara \latin{et~al.}(2010)Enkovaara, Rostgaard, Mortensen, Chen,
  Dulak, Ferrighi, Gavnholt, Glinsvad, Haikola, Hansen, Kristoffersen, Kuisma,
  Larsen, Lehtovaara, Ljungberg, Lopez-Acevedo, Moses, Ojanen, Olsen, Petzold,
  Romero, Stausholm-M{\o}ller, Strange, Tritsaris, Vanin, Walter, Hammer,
  H{\"a}kkinen, Madsen, Nieminen, N{\o}rskov, Puska, Rantala, Schi{\o}tz,
  Thygesen, and Jacobsen]{Enkovaara2010}
Enkovaara,~J.; Rostgaard,~C.; Mortensen,~J.~J.; Chen,~J.; Dulak,~M.;
  Ferrighi,~L.; Gavnholt,~J.; Glinsvad,~C.; Haikola,~V.; Hansen,~H.~A.;
  Kristoffersen,~H.~H.; Kuisma,~M.; Larsen,~A.~H.; Lehtovaara,~L.;
  Ljungberg,~M.; Lopez-Acevedo,~O.; Moses,~P.~G.; Ojanen,~J.; Olsen,~T.;
  Petzold,~V. \latin{et~al.}  \emph{J. Phys. Condens. Matter} \textbf{2010},
  \emph{22}, 253202\relax
\mciteBstWouldAddEndPuncttrue
\mciteSetBstMidEndSepPunct{\mcitedefaultmidpunct}
{\mcitedefaultendpunct}{\mcitedefaultseppunct}\relax
\EndOfBibitem
\bibitem[Hudak \latin{et~al.}(2018)Hudak, Song, Sims, Troparevsky, Humble,
  Pantelides, Snijders, and Lupini]{Hudak18AN}
Hudak,~B.~M.; Song,~J.; Sims,~H.; Troparevsky,~M.~C.; Humble,~T.~S.;
  Pantelides,~S.~T.; Snijders,~P.~C.; Lupini,~A.~R. \emph{ACS Nano}
  \textbf{2018}, \emph{Article ASAP}, doi:10.1021/acsnano.8b02001\relax
\mciteBstWouldAddEndPuncttrue
\mciteSetBstMidEndSepPunct{\mcitedefaultmidpunct}
{\mcitedefaultendpunct}{\mcitedefaultseppunct}\relax
\EndOfBibitem
\bibitem[Ziatdinov \latin{et~al.}(2017)Ziatdinov, Dyck, Maksov, Li, Sang, Xiao,
  Unocic, Vasudevan, Jesse, and Kalinin]{Ziatdinov17AN}
Ziatdinov,~M.; Dyck,~O.; Maksov,~A.; Li,~X.; Sang,~X.; Xiao,~K.; Unocic,~R.~R.;
  Vasudevan,~R.; Jesse,~S.; Kalinin,~S.~V. \emph{ACS Nano} \textbf{2017},
  \emph{11}, 12742--12752\relax
\mciteBstWouldAddEndPuncttrue
\mciteSetBstMidEndSepPunct{\mcitedefaultmidpunct}
{\mcitedefaultendpunct}{\mcitedefaultseppunct}\relax
\EndOfBibitem
\bibitem[Susi, T. (2018)]{opendata}
Susi,~T. \emph{Phaidra Repository} \textbf{2018}, doi:10.25365/phaidra.47\relax
\mciteBstWouldAddEndPuncttrue
\mciteSetBstMidEndSepPunct{\mcitedefaultmidpunct}
{\mcitedefaultendpunct}{\mcitedefaultseppunct}\relax
\EndOfBibitem
\end{mcitethebibliography}

\providecommand{\latin}[1]{#1}
\makeatletter
\providecommand{\doi}
  {\begingroup\let\do\@makeother\dospecials
  \catcode`\{=1 \catcode`\}=2 \doi@aux}
\providecommand{\doi@aux}[1]{\endgroup\texttt{#1}}
\makeatother
\providecommand*\mcitethebibliography{\thebibliography}
\csname @ifundefined\endcsname{endmcitethebibliography}
  {\let\endmcitethebibliography\endthebibliography}{}


%
%
%
%
%
%
%
%
\clearpage

\setcounter{figure}{0}
\renewcommand{\thefigure}{S\arabic{figure}}

\begin{figure}
\includegraphics[width=1.0\textwidth]{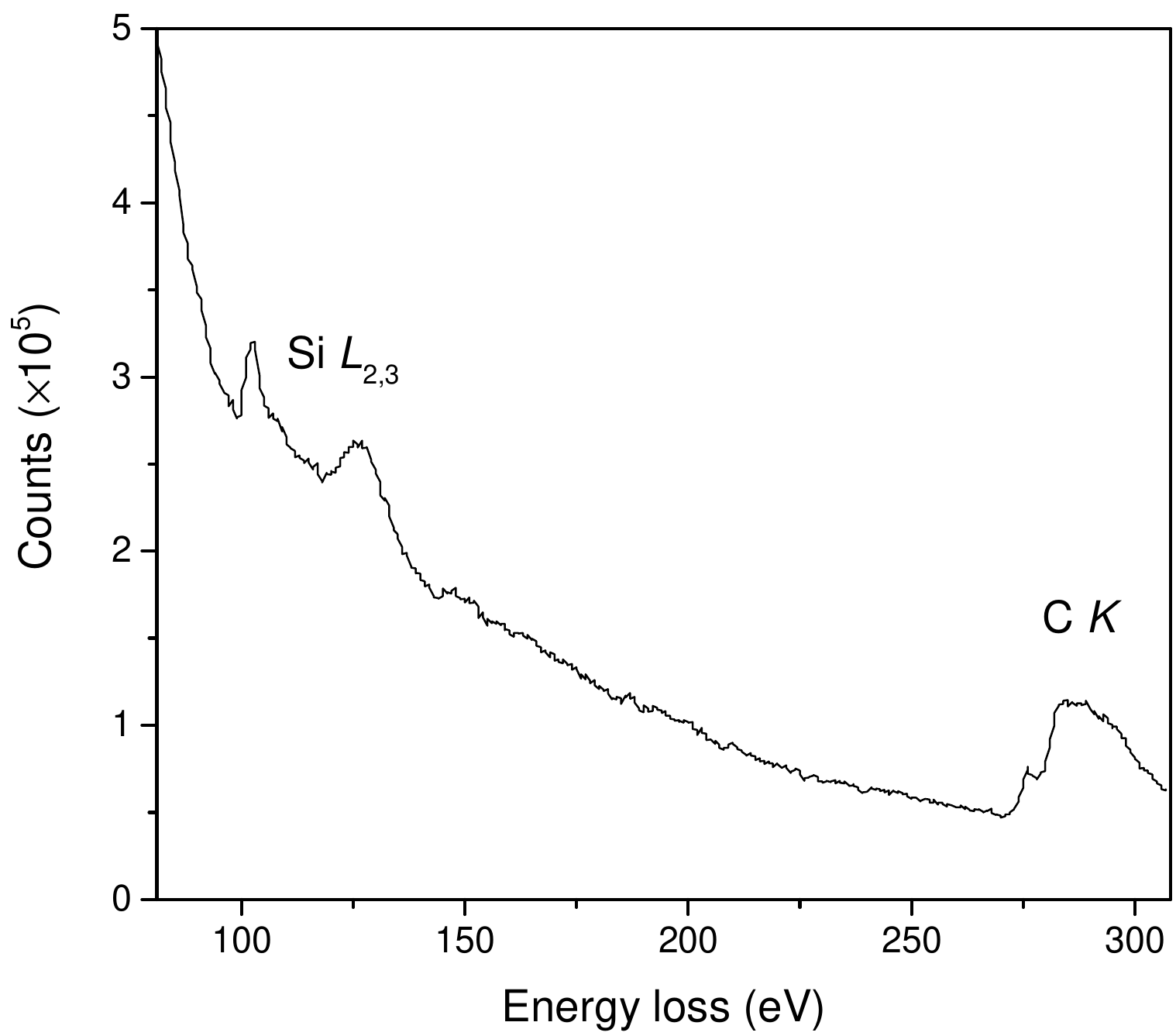}%
\caption{Electron energy loss spectrum measured at an acceleration voltage of 55 kV identifying a single Si impurity by its prominent \textit{L}$_{2,3}$ response.\label{fig:SI_EELS}}
\end{figure}

\begin{figure}
\includegraphics[width=1.0\textwidth]{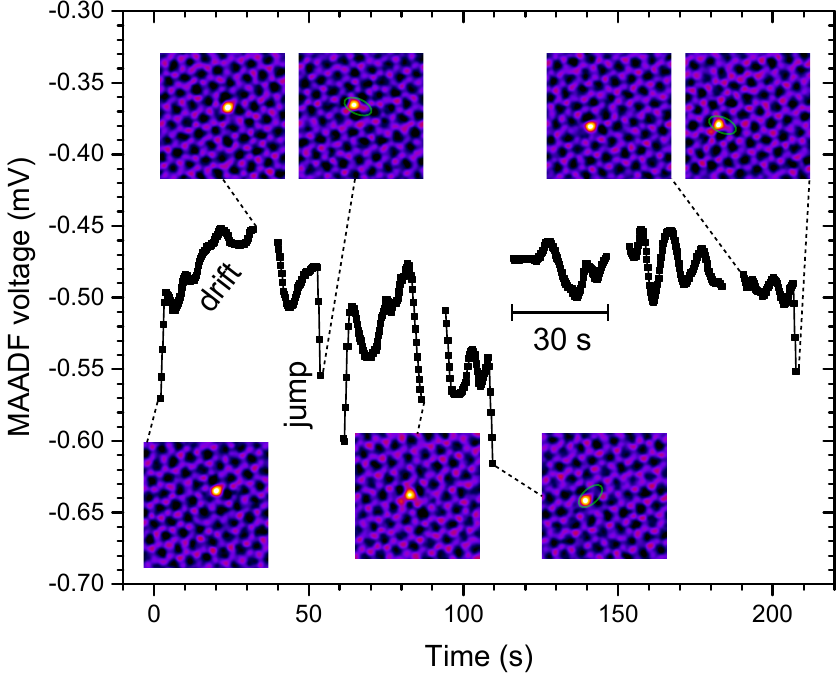}%
\caption{MAADF detector voltage read by a multimeter (each point averaged over 150~ms) and used to trigger new image frames. Jumps are detected as sudden drops in the voltage corresponding to greater scattering intensity due to the Si atom jumping under the beam, whereas gradual drift was accounted for by a 30~s timeout if no jump has been detected. The inset frames show Double Gaussian filtered MAADF images acquired between the feedback periods, with overlaid green ovals highlighting the Si jumps.\label{fig:SI_feedback}}
\end{figure}

\begin{figure}
\includegraphics[width=1.0\textwidth]{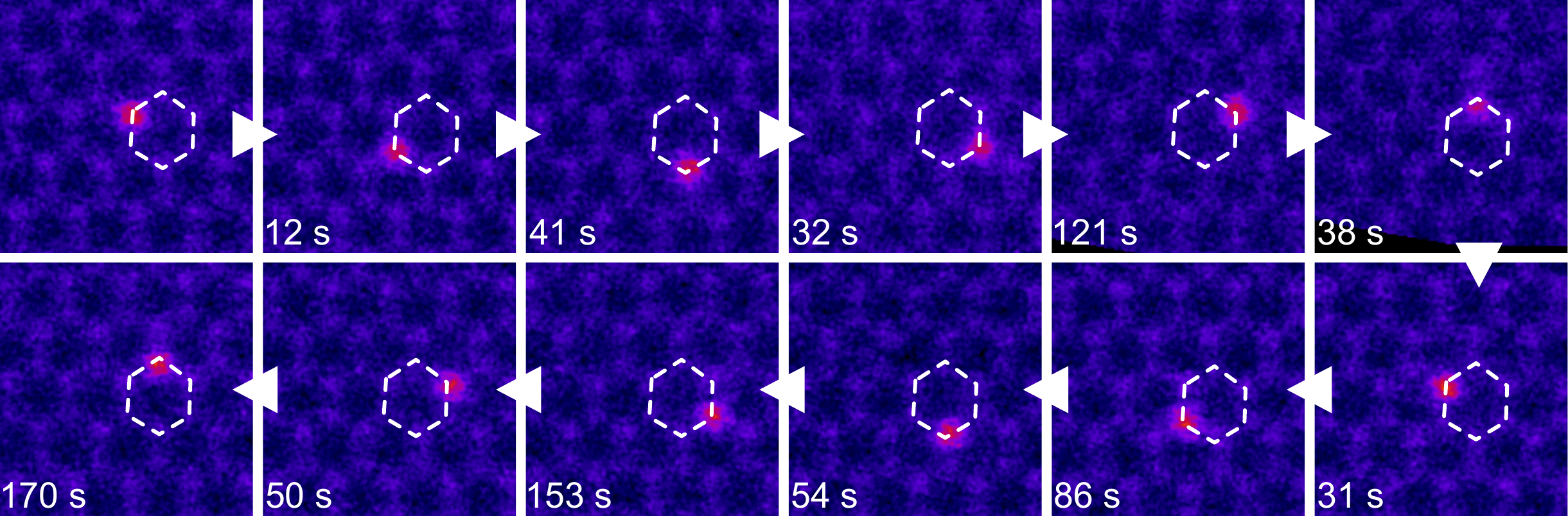}%
\caption{Consecutive STEM/MAADF frames recorded at 55 kV of a Si impurity manipulated around a graphene hexagon. The inset numbers indicate the number of seconds of spot irradiation between frames. (A Double Gaussian filtered, cropped and reordered version of this data appears as the TOC Figure of the main article.)\label{fig:SI_Hex}}
\end{figure}

\begin{figure}
\includegraphics[width=0.7\textwidth]{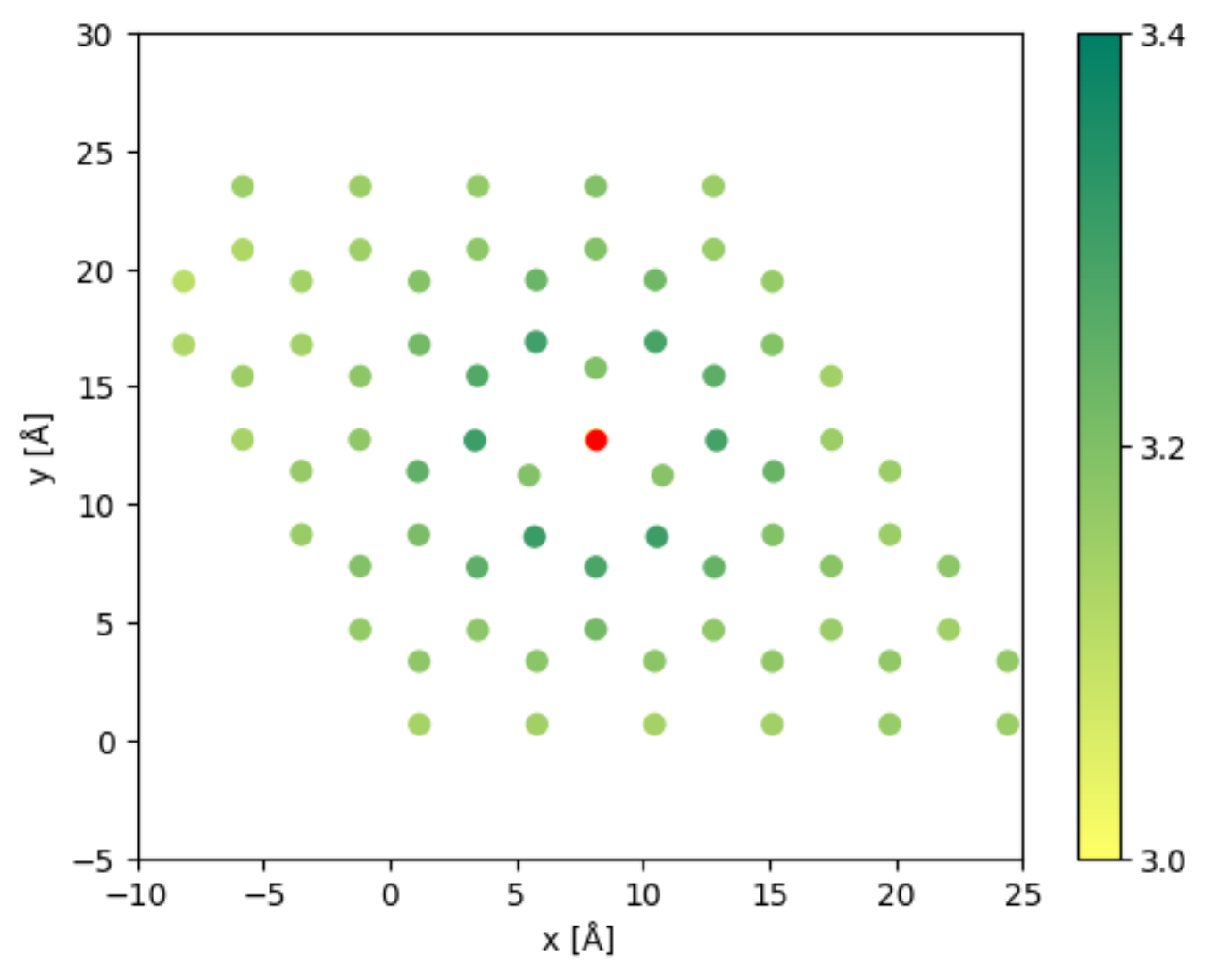}%
\caption{Calculated out-of-plane mean-square velocity for a single Si impurity in graphene, shown as the red dot. Colors of the remaining dots indicate the out-of-plane velocity in units of 10$^5$ m$^2$/s$^2$ on the carbon atoms with the maximum calculated velocity on the next-nearest neighbor carbon atoms to the Si impurity. \label{fig:SI_vz_at_300}}
\end{figure}
 
 
\end{document}